 \documentclass[final,5p,times,twocolumn]{elsarticle}

\usepackage{amssymb}
\usepackage{bm}

\journal{Physics Letter B}

\begin{document}

\begin{frontmatter}



\title{Effect of splitting of the neutron and proton effective mass on nuclear symmetry energy at finite temperature}

 \author[label1,label2,label3]{Li Ou}
 \author[label1,label4]{Zhuxia Li}
 \author[label1]{Yingxun Zhang}
 \author[label2]{Min Liu}
 \address[label1]{China Institute of Atomic Energy, Beijing, 102413, P. R. China}
 \address[label2]{College of Physics and Technology, Guangxi Normal University, Guilin, 541004, P. R. China}
 \address[label3]{only.ouli@gmail.com}
 \address[label4]{lizwux@ciae.ac.cn}

\begin{abstract}
We present the temperature and density dependence of symmetry
energy for nuclear matter at finite temperature based on the
approach of the thermodynamics with Skyrme energy density
functional. We first classify the Skyrme interactions into 7
groups according to the range of neutron and proton effective mass
in neutron matter limit(99.99 per cent neutron in the matter). We
find that there is obvious correlation between the temperature
dependence of the symmetry energy and the splitting of the neutron
and proton effective mass. For some Skyrme interactions with
$m^{*}_{n}>m^{*}_{p}$ and strong splitting of the neutron and
proton effective mass in asymmetric nuclear matter, a transition
of the temperature dependence of symmetry energy from decreasing
with temperature at low densities to increasing with temperature
at high densities appears.
For other Skyrme interactions, we do not observe such phenomenon.
Our study show that the symmetry energy in hot asymmetric matter
not only depends on symmetry potential part but also on the
splitting of the neutron and proton effective mass to a certain
extent.
\end{abstract}


\end{frontmatter}

The symmetry energy for nuclear matter $E_{\rm{sym}}$
is very important for
understanding not only the structure of nuclei far from $\beta$
stability line, the dynamics and many interesting phenomena in
reactions of isospin asymmetric nuclei but also many critical
issues in astrophysics\cite{Liba08}.
Recent research on the symmetry energy for cold nuclear matter has
already obtained a great progress and some constraints on the
density dependence of the symmetry energy at subnormal densities
have been set \cite{Liutx07,Famiano06,Liba05,Tsang09,Cent09}. But
the density dependence of the symmetry energy at high densities
extracted from the ratio of $\pi^{-}/\pi^{+}$ and other
observables in heavy ion collisions at intermediate high energies
are extremely model dependent and divergent\cite{Xiao09,Trau09}.
Moreover, for heavy ion collisions at intermediate high energies,
the system should be heated and therefore in addition to the
density dependence, the temperature dependence of the symmetry
energy is also of fundamental importance for heavy ion collisions
with nuclei away from $\beta$-stability line as well as for
astrophysics. While, how the symmetry energy depending on
temperature is even unclear\cite{Xuj08,Moustakidis07,Liba06} and
only few information on the symmetry energy at very low density
and very low temperature was extracted from experimental
measurement in heavy ion collisions up to now\cite{Kowalski07}

The symmetry energy for hot nuclear matter can be expressed as
\begin{eqnarray}\label{Esym}
E_{\rm{sym}}(\rho, T)=E(\rho, T,\delta=1)-E(\rho, T,\delta=0).
\end{eqnarray}
$E(\rho, T,\delta)$ is the energy per nucleon for nuclear matter
with density $\rho$, temperature $T$, isospin asymmetry
$\delta$=$\frac{\rho_{n}-\rho_{p}}{\rho}$, where $\rho_{n}$,
$\rho_{p}$, and $\rho$ are the neutron, proton and nucleon density
of the system, respectively. It gives an estimation of the energy
cost to convert all protons in symmetric nuclear matter(SM) to
neutrons at fixed temperature $T$ and density $\rho$.

In addition to the symmetry energy, the splitting of the neutron
and proton effective mass (EFM) is also a hot topic in the study
of the properties of asymmetry nuclear
matter\cite{Bethe90,Farine01}. The neutron and proton EFM
splitting predicted by different effective interactions are rather
different\cite{Xuj08,Dalen05,Rizzo04,Persram02,Pandharipande92,Liba04}.
Both the symmetry energy and the neutron and proton EFM splitting
closely relate to the isospin dependence of the nuclear
interactions. Then one would naturally ask whether the splitting
of the neutron and proton EFM influences the temperature and
density dependence of the symmetry energy? In this letter we will
first study the symmetry energy at finite temperature and then
investigate the correlation between the EFM splitting and the
density and temperature dependence of symmetry energy. The
approach of thermodynamics with Skyrme energy density functional
is applied in this work.

The Skyrme energy density functional reads
\begin{eqnarray}\label{Ham}
H&=&\textstyle{\hbar^{2} \over {2m}}
[\tau_{p}(\bm{r})+\tau_{n}(\bm{r})]\nonumber \\
&&+\textstyle{1 \over
4}t_{0}[(2+x_{0})\rho^{2}-(2x_{0}+1)(\rho_{p}^{2}+\rho_{n}^{2})]
\nonumber \\
&& +\textstyle{1 \over
24}t_{3}\rho^{\alpha}[(2+x_{3})\rho^{2}-(2x_{3}+1)(\rho_{p}^{2}+\rho_{n}^{2})]
\nonumber \\
&& +\textstyle{1 \over 8}[t_{1}(2+x_{1})+t_{2}(2+x_{2})]\tau \rho\nonumber \\
&& +\textstyle{1 \over
8}[t_{2}(2x_{2}+1)-t_{1}(2x_{1}+1)](\tau_{p}\rho_{p}+\tau_{n}\rho_{n}),
\end{eqnarray}
where $\rho=\rho_{n}+\rho_{p}$, and $\tau=\tau_{n}+\tau_{p}$. The
$\rho_{q}$ and $\tau_{q}$ ($q$ denotes protons or neutrons) are
computed by
\begin{eqnarray}\label{rhoq}
\rho_{q}=2\int_{0}^{\infty} n_{q}(p) \textstyle{4\pi p^{2}\over
\hbar^{3} } d p,
\end{eqnarray}
and
\begin{eqnarray}\label{tauq}
\tau_{q}=2\int_{0}^{\infty} n_{q}(p) \textstyle{4\pi p^{2}\over
\hbar^{3} } \textstyle{p^{2} \over \hbar^{2}}d p.
\end{eqnarray}
The occupation number distribution for species q, $n_{q}(p)$,
obeys the Fermi-Dirac distribution function, which reads
\begin{eqnarray}\label{FDF}
n_{q}(p)=\frac{1}{1+\exp[\beta(\varepsilon_{q}-\mu_{q})]}.
\end{eqnarray}
 $\varepsilon_{q}=\frac{p^{2}}{2 m^{*}_{q}}+U_{q}$ is the
single particle energy for species q. The EFM $m_{q}^{*}$ for
species q is expressed as
\begin{eqnarray}\label{Meff}
\textstyle{\hbar^{2} \over 2 m_{q}^{*}}&=&\textstyle{\hbar^{2} \over
2 m_{q}}+\textstyle{1 \over
8}[t_{1}(2+x_{1})+t_{2}(2+x_{2})]\rho \nonumber \\
&&+\textstyle{1 \over 8}[t_{2}(2x_{2}+1)-t_{1}(2x_{1}+1)]\rho_{q}.
\end{eqnarray}
The mean field $U_{q}$ for species q reads
\begin{eqnarray}\label{MF}
U_{q}(r)&=&\textstyle{1 \over 2}
t_{0}[(2+x_{0})\rho-(1+2x_{0})\rho_{q}]\nonumber \\
&&+\textstyle{1 \over 24} t_{3}[(2+x_{3})(2+\alpha)\rho^{\alpha+1}\nonumber \\
&&-(1+2x_{3})[2\rho^{\alpha}\rho_{q}+\alpha\rho^{\alpha-1}(\rho_{n}^{2}+\rho_{p}^{2})]]
\nonumber \\
&& +\textstyle{1 \over 8}[t_{1}(2+x_{1})+t_{2}(2+x_{2})]\tau \nonumber \\
&& +\textstyle{1 \over 8}[t_{2}(2x_{2}+1)-t_{1}(2x_{1}+1)]\tau_{q}.
\end{eqnarray}
Introducing effective chemical potential
$\mu^{'}_{q}=\mu_{q}-U_{q}$, then
\begin{eqnarray}\label{FDFe}
n_{q}(p)=\frac{1}{1+\exp[\beta(p^{2}/2m_{q}^{*}-\mu^{'}_{q})]}.
\end{eqnarray}
By solving equation (\ref{rhoq}) and (\ref{FDFe}) with EFM
depending on density iteratively, for any pair of $\mu^{'}_{n}$
and $\mu^{'}_{p}$ we can obtain the proton and neutron density
$\rho_{p}$ and $\rho_{n}$ at temperature $T$. Then, the energy per
nucleon in neutron matter(NM) and SM can be calculated by using
equation (\ref{tauq}) and (\ref{MF}) and finally symmetry energy
at finite temperature can be calculated.

  \begin{table*}
\tabcolsep 0pt \caption{\label{Table1} The ranges of the neutron
and proton EFMes, $m^{*}/m$, in SM and NM calculated with various
Skyrme interactions.}
 \vspace*{-12pt}
\begin{center}
\def\temptablewidth{1.\textwidth}
\rule{\temptablewidth}{1pt}
\begin{tabular*}{\temptablewidth}{@{\extracolsep{\fill}}ccc}
 & \hspace{-5cm}group\hspace{3cm}SM & NM(99.99 percent neutrons)  \\
\hline
 I  $m_{p}^{*}=m_{n}^{*}>$1 & \parbox{6.5cm}{BSk1$\sim$3, MSk2, MSk4$\sim$6,
 SVII, SKXce, $v$~series} &\parbox{6.5cm}{ BSk1, MSk2,
MSk4$\sim$6, v105}\\
\hline
 II  $m_{p}^{*}=m_{n}^{*}=$1 & \parbox{6.5cm}{ MSk1, MSk3, SkP,
 SkSC1$\sim$4, SkT1$\sim$6} &\parbox{6.5cm}{ MSk1, MSk3,
 SkSC1$\sim$4, SkT1$\sim$6}\\
\hline
 III  $m_{p}^{*}=m_{n}^{*}<$1 &\parbox{6.5cm}{ BSK4$\sim$17, Es, FitB, Gs, RATP, Rs,
 SGI, SGII,SI$\sim$VI, SIII$^{*}$, SKRA, SkI1$\sim$6,
 SkM1, SkM, SkM$^{*}$, SkMP, SkO, SkT7$\sim$9,
 SKX, SKXm, Skz~series, SLy~series, Zs} & \parbox{6.5cm}{ SkT8, SkT9} \\
\hline
IV  $m_{p}^{*}<1<m_{n}^{*}$ & ~&
 \parbox{6.5cm}{ BSK2$\sim$5, BSK10$\sim$13, Es, FitB, SI, SkP, SKX, SKXce, SKXm, Skz-1,
 Skz0, SVI, SVII,  $v$070, $v$075, $v$080, $v$090, $v$100}\\
 \hline
 V  $m_{p}^{*}<m_{n}^{*}<$1 & ~& \parbox{6.5cm}{
 BSK14$\sim$17, Gs, RATP, Rs, SGI, SGII, SII$\sim$SV, SIII$^{*}$,
 SKRA, SkM, SkM1, SkMP, SkM$^{*}$, SkO,
 SkT7, Skz1, Skz2, Zs}\\
\hline VI  $1>m_{p}^{*}>m_{n}^{*}$ & ~&
 \parbox{6.5cm}{ BSK6$\sim$9, SkI1$\sim$6, Skz3, Skz4, SLy-series}\\

\hline
 VII  $m_{p}^{*}>m_{n}^{*}>1$ & ~&
 $v$110\\
\end{tabular*}
{\rule{\temptablewidth}{1pt}}
\end{center}
\end{table*}

Since the nucleon EFM is explicitly involved in the iteration
process we first make a general survey of the density dependence
of the neutron, proton EFM for various Skyrme interactions. 94
Skyrme interactions are tested and they can be divided into 7
groups according to their corresponding ranges of the neutron and
proton EFMes varying with density. Table~\ref{Table1} presents the
groups of Skyrme interactions and the corresponding ranges of
$m^{*}/m$ in SM and NM. The $m$ is the bare nucleon mass. In this
work, the NM is always referred to a nuclear matter limit with
99.99 percent neutrons in the matter. For convenience, the EFM
$m^{*}$ discussed in the following text always refer to the ratio
$m^{*}/m$ (the scaled effective mass), which should not be
confused with the EFM $m_{q}^{*}$ appeared in expression (6) and
(8). In SM, the neutron and proton EFM are equal and all Skyrme
interactions are simply in the groups: (I) $m^{*}>1$ and it
increases with density increasing; (II) $m^{*}=1$ and it is
independent on density; and (III) $m^{*}<1$ and it decreases with
density increasing. Most of Skyrme interactions belong to the
group (III). For asymmetric nuclear matter, the situation becomes
more complex because of neutron-proton EFM splitting. The
splitting of the neutron and proton EFM in asymmetric matter is
currently not known empiricaly\cite{Xuj08}. Theoretical results on
the splitting of the neutron and proton EFM are highly
controversial among different approaches and different effective
interactions. One can see from Table I that only few Skyrme
interactions belong to groups I,II, and III and most of Skyrme
interactions belong to groups IV to VII with splitting of neutron
and proton EFM. Large part of them belong to groups IV and V
($m^{*}_{n}>m^{*}_{p}$), others belong to groups VI and VII
($m^{*}_{n}<m^{*}_{p}$).
\begin{figure}
\includegraphics[width=0.4\textwidth]{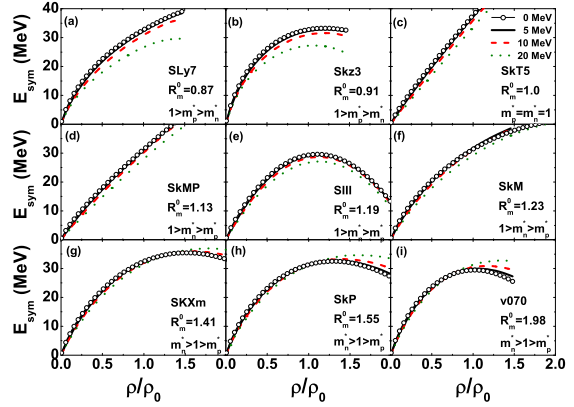}
\caption{(color online) Density dependence of symmetry energy at
$T$=0, 5, 10, 20 MeV calculated with different Skyrme
interactions. The ranges of corresponding neutron and proton EFM
and the ratio between the neutron EFMes in NM and SM at saturation
density are also presented in each
sub-figure.}{\label{EsymSkyrme}}
\end{figure}

Now let us come to study the density and temperature dependence of
the symmetry energy in hot nuclear matter. We show in Fig.
\ref{EsymSkyrme} the symmetry energies in nuclear matter at
temperatures $T$= 0, 5, 10, 20 MeV calculated with different
Skyrme interactions, respectively, as typical examples. The 9
subfigures in Fig.\ref{EsymSkyrme} is ordered according to the
magnitude of $R_{m}^{0}$ for the Skyrme interaction applied. Here
we introduce the ratio $R_{m}$= $\frac{m^{*}_{1}}{m^{*}_{0}}$,
where the subscripts 0 and 1 indicate the isospin asymmetry
$\delta$ =0 (for SM) and 1 (for NM), respectively, to characterize
the strongness of the splitting of the neutron and proton EFM for
the Skyrme interaction applied. The quantity $R^{0}_{m}$ presented
in each subfigure is the value of $R_{m}$ at normal density.
Obviously, the $R_{m}$ is proportional to the neutron and proton
EFM splitting in asymmetric matter. The values of $R^{0}_{m}$ for
Skyrme interactions in groups IV and V are always larger than 1
and those in groups VI and VII are smaller than 1. As is known
that the symmetry energies for cold nuclear matter calculated from
different Skyrme interactions are largely divergent especially at
high densities. We can also classify the Skyrme interactions into
two groups A and B according to the trend of the density
dependence of the symmetry energy in cold matter. For the group A,
namely the Skyrme interactions like SLy7, SkT5, SkMP, and SkM,
etc., nuclear symmetry energy increases monotonously with density
and for the group B, namely the Skyrme interactions like Skz3,
SIII, SKXm, SkP, and $v$070, the nuclear symmetry energy first
increases with density until certain density then it bends down as
density further increases and finally becomes negative. The group
A corresponds to the group I and the group B corresponds to group
II and III in ref.\cite{St03}, respectively. It is seen from Fig.1
that there is no obvious correlation between the trend of the
density dependence of the symmetry energy in cold matter and the
magnitude of the $R^{0}_{m}$ of the corresponding Skyrme
interactions.
As the temperature dependence of the symmetry energy is concerned,
one sees that the nuclear symmetry energy decreases with
temperature increase in all densities for Skyrme interactions with
small $R^{0}_{M}$, such as SLy7, Skz3,Skt5,SkMP,etc., in
consistency with the results given in \cite{Xuj08}. But for SkM,
SKXm, SkP, and $v$070 with large $R^{0}_{m}$ the symmetry energy
decreases with temperature increasing at low density and increases
with temperature when the density is higher than a certain
density. We call this phenomenon the transition of temperature
dependence of the symmetry energy(TrTDSE). From Table \ref{Table1}
one can find that Skyrme interactions such as SkM, SKXm, SkP, and
$v$070 for which the TrTDSE phenomenon appears, all belong to
groups IV and V satisfying $m^{*}_{n}
> m^{*}_{p}$. The density for the onset of the TrTDSE depends on
the magnitude of the $R_{m}^{0}$ of the corresponding Skyrme
interactions. The larger $R_{m}^{0}$ is, the lower density is for
the onset of the TrTDSE. For the case of SIII, we find that the
symmetry energy at finite temperature is very close to that of
cold matter up to $\rho \leq 4\rho_{0}$ where our calculation
stops. For Skyrme interactions belonging to the groups I-III and
VI,VII (not satisfying $m^{*}_{n}> m^{*}_{p}$), the TrTDSE
phenomenon can not happen.

\begin{figure}
\includegraphics[width=0.4\textwidth]{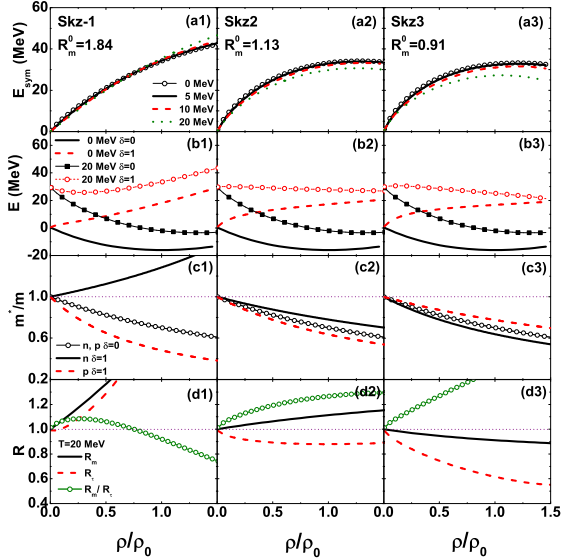}
\caption{(color online) The symmetry energy at $T$=0, 5,10,20 MeV,
the energy per nucleon at $T$=0 and 20 MeV , the neutron and
proton EFMes , the $R_{m}$, $R_{\tau}$, and $R_{m}/R_{\tau}$ at
$T$=20 MeV in NM and SM as function of density calculated with
Skz-1, Skz2 and Skz3 interactions, respectively.}{\label{EsymSkz}}
\end{figure}

In order to explore the condition of TrTDSE in hot matter more
quantitatively we study the difference between the symmetry energy
in a matter at temperature T and that in a cold matter, which
reads as
\begin{eqnarray}\label{DEsym}
\Delta E_{\rm{sym}}(\rho,T)&=&E_{\rm{sym}}(\rho,T)-E_{\rm{sym}}(\rho,0)\nonumber \\
&=&(\textstyle{\hbar^{2} \over
2m\rho}+b_{3}+b_{4})(\tau^{T}_{1}-\tau^{0}_{1})(1-\textstyle{R_{m}
\over R_{\tau}}),
\end{eqnarray}
with $b_{3}=\textstyle{1 \over 8}[t_{1}(2+x_{1})+t_{2}(2+x_{2})]$
and $b_{4}=\textstyle{1 \over
8}[t_{2}(1+2x_{2})-t_{1}(1+2x_{1})]$. It is easy to extend
equation (9) to $E_{\rm{sym}}(\rho,\Delta T)$ with $\Delta T
=T_{2}-T_{1}$. The $R_{\tau}$ in (9) is the ratio between the
kinetic energy density increasing from $T=0$ to $T=T$ in NM and
that in SM, $R_{\tau}=\textstyle{\tau^{T}_{1}-\tau^{0}_{1} \over
\tau^{T}_{0}-\tau^{0}_{0}}$, where the subscripts 0 and 1 indicate
the isospin asymmetry $\delta$ =0 and 1, respectively. The first
factor in the expression (\ref{DEsym}) depends on the Skyrme
interaction parameters which is always positive up to
$\rho=2\rho_{0}$ for the 94 Skyrme interactions involved in this
work( in fact for some Skyrme interactions this density can be
much higher than $2\rho_{0}$). The second factor is always
positive. Consequently the sign of $\Delta E_{\rm{sym}}(\rho,T)$
is determined by the third factor. If $R_{m}/R_{\tau}>$1 the
symmetry energy decreases with temperature, otherwise it increases
with temperature. The splitting of the neutron and proton EFM
influences the $\Delta E_{\rm{sym}}(\rho,T)$ explicitly through
$R_{m}$ and indirectly through $R_{\tau}$. Now let us investigate
how the $R_{\tau}$ is eventually also influenced by the EFM
splitting. Before coming to this point, we show in Fig.
\ref{EsymSkz} the systematic calculated results
with Skz-1, Skz2, and Skz3, respectively. We notice that the
predictions from Skz-series Skyrme interactions are the same for
the isoscalar part but different for the isovector part of the
equation of state and moreover the Skz-1, Skz2, and Skz3 belong to
groups IV, V, VI, respectively. Therefore they are very suitable
for exploring the correlation between the temperature dependence
of the symmetry energy and the nucleon EFM splitting. In Fig.
\ref{EsymSkz}, the top panel shows the density dependence of
symmetry energies at $T$=0, 5, 10, 20 MeV, respectively; the
second panel shows the energies per nucleon in NM and SM at $T$=0
and 20 MeV, where the energies per nucleon in SM are the same for
three interactions because they have the same isoscalar part; The
third panel shows the nucleon(neutron) EFMes in SM and NM, and the
bottom panel shows the density dependence of $R_{m}$, $R_{\tau}$,
and $R_{m}/R_{\tau}$ at $T$=20 MeV, respectively. The results
calculated with other Skz-series interactions are similar to the
one of these three Skz interactions. We first investigate the
calculation results with Skz-1, the strong neutron and proton EFM
splitting case with $R_{m}^{0}$=1.84, one sees from Fig.2(c1) that
the neutron EFM in NM is larger than 1 and increases with density
quickly, the proton EFM is smaller than 1 and decreases with
density even faster, and the nucleon EFM in SM is just between
them and decreases with density, which leads the $R_{m}$ to
increase with density quickly as shown in Fig.2(d1). From
Fig.2(b1), one sees that the energy per nucleon in NM increases
with density and that in SM decreases with density for both $T$=20
and $T$=0 MeV cases but the increasing slopes in NM and decreasing
slopes in SM with density are different for $T$=20 and $T$=0 MeV.
As a result, the symmetry energy at $T$=20 MeV becomes higher than
that at $T$=0 MeV when $\rho>$0.75$\rho_{0}$(Fig.2(a1). It is more
evident from Fig.2(d1). For T=20 MeV case, the $R_{m}/R_{\tau}$
becomes smaller than one when the density exceeds about
0.75$\rho_{0}$ and therefore the symmetry energy to be larger than
that in T=0 MeV case from Eq.(9). In the middle and right columns,
we show the results calculated with Skz2 and Skz3. For Skz2(
$m^{*}_{n}>m^{*}_{p}$ with $R_{m}^{0}$=1.13) and for Skz3(
$m^{*}_{n}< m^{*}_{p}$ with $R_{m}^{0}$=0.91), the splitting of
the neutron and proton EFM is small and both neutron and proton
EFM are close to 1. Accordingly, the $R_{m}/R_{\tau}$ becomes
larger than 1 at T=20 MeV for both Skz2 and Skz3 cases and thus
the symmetry energies for those two cases are all decreasing with
temperature in whole densities. The extent of decreasing depends
on the magnitude of $R_{m}/R_{\tau}$. The calculation results for
$v$-series Skyrme interactions have similar behavior as those of
Skz-series.
\begin{figure}
\includegraphics[width=0.4\textwidth]{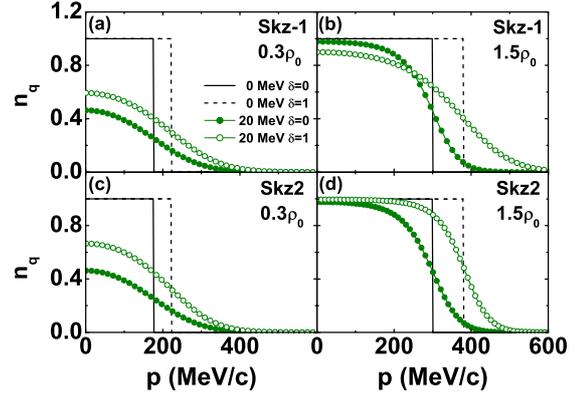}
\caption{(color online) The neutron(proton) occupation number
distribution functions in NM and SM at $T$=0, 20 MeV calculated
with Skz-1 and Skz2, respectively.}{\label{FDD}}
\end{figure}

To understand what factors will influence the $R_{\tau}$, let us
start from the expression (\ref{FDF}) or (\ref{FDFe}) for
$n_{q}(p)$. For cold matter, the $n_{q}(p)$ is a step-function,
the momenta of all particles are within the Fermi momentum sphere
with $k_{nf}=k_{pf}=(3\pi^{2}\rho/2)^{1/3}$ for SM and
$k_{nf}=(3\pi^{2}\rho)^{1/3}$ for NM. For hot matter we know from
expressions (\ref{FDF}) or (\ref{FDFe}) that the deviation of
$n_{q}(p)$ from a step-function depends on the density (related to
the chemical potential term in (\ref{FDF}) and (\ref{FDFe})) and
nucleon EFM in the matter(related to the kinetic energy of nucleon
in (\ref{FDFe})) when temperature is fixed. More explicitly,the
deviation of $n_{q}(p)$ for hot matter from a step-function is
larger for a system with lower density and larger nucleon EFM.
In Fig. \ref{FDD} the left panel shows the $n_{q}(p)$ at
$\rho=0.3\rho_{0}$ and the right panel shows the $n_{q}(p)$ at
$\rho=1.5\rho_{0}$ at T=0, 20 MeV calculated with Skz-1 and Skz2,
respectively. The solid lines in the figure denote $n_{q}(p)$ for
neutrons(protons) in SM and dashed lines denote $n_{n}(p)$ for
neutrons in NM at T=0. The excessive kinetic energy needed for
conversion of all protons in SM into neutrons in NM comes from the
excessive neutrons in $n_{q}(p)$ in NM as compared with that in
SM, which is one of the origin of the symmetry energy for cold
matter. For the matter at finite temperature, the calculated
results of $n_{q}(p)$ are shown by the lines with symbols in the
figure. The calculation results are just as expected. From
Fig.3(a) and Fig.3(c) for low density $\rho=0.3\rho_{0}$ case we
find that the $n_{q}(p)$ at T=20 MeV for both $\delta=0$ and
$\delta=1$ deviates from step-function largely but there is no
obvious difference between the results calculated with Skz-1 and
Skz2 because of low matter density. For high density
$\rho=1.5\rho_{0}$ case, the results of $n_{q}(p)$ (right panel)
in SM ($\delta$=0) at $T$=20 MeV calculated with Skz-1 and Skz2
are the same because of the same iso-scalar part of two
interactions and both deviate from step-function not largely.
While, the results of $n_{q}(p)$ in NM ($\delta$=1) calculated
with Skz-1 and Skz2 become much different. For the case with
Skz2(Fig.3(d)), the deviation of the nucleon $n_{q}(p)$ in NM from
a step-function is almost the same or even weaker than that in SM
and the ratio $R_{\tau}$ becomes smaller than 1(see Fig.2(d2)),
which leads the $R_{m}$/$R_{\tau}$ at T=20 MeV to be larger than 1
because $R_{m}>1$ and consequently the symmetry energy at T=20 MeV
becomes smaller than that at $T$=0. While for the case with Skz-1,
the nucleon $n_{n}(p)$ at $T$=20 MeV in NM deviates from that at
$T$=0 MeV largely (Fig.3(b)) and its very long tail means that a
large portion of neutrons inside the Fermi momentum sphere are
excited to the outside of the Fermi momentum sphere and quite
large part of them have very large momenta. And consequently the
kinetic energy increases largely. Thus, the $R_{\tau}$ becomes
much large at high density and even larger than the $R_{m}$ which
is larger than one. Finally, $R_{m}/R_{\tau}$ at high densities
becomes smaller than 1 and the symmetry energy at high density at
$T$=20 MeV is enhanced as compared with cold matter. From above
discussion, we find that the splitting of the neutron and proton
EFM influences the $R_{\tau}$ through the difference between the
$n_{q}(p)$ in SM and that in NM. This effect on $R_{\tau}$ can
even be amplified as compared with that on $R_{m}$ for some Skyrme
interactions. Thus, ultimately, the splitting of the neutron and
proton EFM influences the trend of the density dependence of
symmetry energy in hot matter. For some Skyrme interactions like
Skz-1 which has strong splitting of the neutron and proton EFM and
$m_{n}^{*}>m_{p}^{*}$, the onset of the TrTDSE at a certain
density, usually at supernormal density, leads the symmetry energy
to be stiffer in hot matter than that in cold matter. For others,
the splitting of the neutron and proton EFM leads the symmetry
energy to be more soft in hot matter than that in cold matter.
Generally, if the Skyrme interaction belongs to group IV and V
with its $R_{m}^{0}$ being much larger than 1, the TrTDSE will
occur at certain density. The density for the onset of the TrTDSE
depends on the magnitude of the splitting of the neutron and
proton EFM in asymmetric matter. Concerning experimentally probing
the TrTDSE or the temperature and density dependence of the
symmetry energy, it requires to simultaneously measure the
observables such as the transverse momentum spectrum of
protons(pions,or other light charged particles) to determine the
temperature \cite{Xiao} and those to be sensitive to symmetry
energy at high densities such as the ratio between the yields of
different charged pions, $\pi^{-}/\pi^{+}$, to extract the
symmetry potential at high densities in heavy ion collisions with
isospin asymmetric nuclei at energies of hundreds to around 1 GeV
per nucleon. At present, highly contradictive results about the
symmetry energy at high density extracted from experiments may
have some relation with the unclear temperature dependence of the
symmetry energy.

In summary, we have investigated the temperature and density
dependence of the symmetry energy in hot nuclear matter by means
of the approach of the thermodynamics with Skyrme energy density
functional. We first classify the Skyrme interactions into 7
groups according to the range of the calculated neutron and proton
EFM varying with density in the NM limit. Then we study the
temperature dependence of symmetry energy for different Skyrme
interactions. We find that for some Skyrme interactions the
symmetry energy decreases with temperature at whole density region
and it means that the symmetry energy in hot matter is softer than
that in cold matter. But for Skyrme interactions belonging to
groups IV and V (for these two groups $m^{*}_{n}>m^{*}_{p}$), the
TrTDSE phenomenon may occur at certain density, which makes the
symmetry energy at high density to be stiffer. In order to explore
the correlation between the neutron and proton EFM splitting and
the temperature dependence of the symmetry energy in hot nuclear
matter we introduce the ratio $R_{m}$ between the nucleon EFM in
NM and that in SM, which is directly related with the neutron and
proton EFM splitting in asymmetric matter and the ratio $R_{\tau}$
between the kinetic energy density increasing from $T$=0 to $T=T$
in NM and that in SM. The effect of the effective mass splitting
on $R_{\tau}$ is indirect but for some Skyrme interactions it can
be amplified as compared with that on $R_{m}$. Only for those
Skyrme interactions with $R_{m}/R_{\tau}<$1 in group IV and V the
TrTDSE phenomenon can occur at high densities and it ultimately
makes the trend of the density dependence of symmetry energy to be
more stiff in hot matter. The onset density of TrTDSE depends on
the magnitude of $R_{m}^{0}$. The larger $R_{m}^{0}$ is, the lower
onset density is. Finally, we would stress that the symmetry
energy in hot nuclear matter not only depends on symmetry
potential part but also on the splitting of the neutron and proton
EFM in kinetic part to a certain extent. Therefore, the study of
the temperature dependence of the symmetry energy in addition to
the density dependence is a matter of significance for extracting
the symmetry potential and also the splitting of the neutron and
proton EFM in asymmetry nuclear matter.

This work has been supported by the National Natural Science
Foundation of China under Grant Nos 10875031, 10975095,
10905021,1097923,10847004, 11005022, 11075215 and the National
Basic research program of China No. 2007CB209900.


\end{document}